**SUMO Substrates and Sites Prediction**

**Combining Pattern Recognition and Phylogenetic Conservation**

Yu Xue[1], Fengfeng Zhou[2], Hualei Lu[1], Guoliang Chen[2] and Xuebiao Yao[*1, 3]

[1]Laboratory of Cell Dynamics, University of Science & Technology of China, Hefei, CHNA 230027;

[2]National High Performance Computing Center at Hefei, Department of Computer Science and Technology, University of Science and Technology of China, Hefei, Anhui, P.R.China, 230027;

[3]Department of Molecular & Cell Biology, University of California, Berkeley, CA 94720, USA

***Running title*:** SUMO Substrates and Sites Prediction

¶ ***All correspondence addressed to:***

Xuebiao Yao

Email: xbyao@berkeley.edu

[*] To whom correspondence should be addressed

The first two authors contributed equally to this work.

## Abstract

***Motivation:*** Small Ubiquitin-related modifier (SUMO) proteins are widely expressed in eukaryotic cells, which are reversibly coupled to their substrates by motif recognition, called sumoylation. Two interesting questions are 1) how many potential SUMO substrates may be included in mammalian proteomes, such as human and mouse, 2) and given a SUMO substrate, can we recognize its sumoylation sites? To answer these two questions, previous prediction systems of SUMO substrates mainly adopted the pattern recognition methods, which could get high sensitivity with relatively too many potential false positives. So we use phylogenetic conservation between mouse and human to reduce the number of potential false positives.

***Results:*** We first use two major patterns of a potential SUMO substrate: an NLS (Nuclear Localization Signal) and a consensus motif ψ-K-X-E, where ψ meant a hydrophobic amino acid. So we followed a simple rule to predict the SUMO substrates: the sub-cellular localization of a given protein must be predicted as nuclear by PSORT II, and the protein must have at least one consensus motif ψ(A, F, I, L, M, V, W)-K-X-E. After the above methods, there are still too many predicted positives. To eliminate the potential false positives, we used the orthology information between mouse and human, and followed the rule that at least one consensus motif should be at the same position after sequence alignment of the ortholog pair. We got 2,683 potential SUMO substrates in both mouse and human, with 58 out of 79 known SUMO substrates in human predicted correctly. For the sumoylation sites prediction, our method got nearly the same sensitivity as the existed tool SUMOplot, with 42 against 42 (high probability) or 44 (all) true positives of 54 known sumoylation sites respectively. But our method outperformed SUMOplot significantly by specificity, with 74 against 152 (high probability) or 324 (all) predicted sumoylation sites respectively. So our method greatly reduced the number of potential false positives while still kept a satisfying sensitivity.

***Availability:*** The software SSP (SUMO Substrate and Site Prediction) 1.0 for Windows system, written by Delphi, is available from http://973-proteinweb.ustc.edu.cn/sumo/.

***Contact*:** yxue@mail.ustc.edu.cn, ffzhou@ustc.edu

# Introduction

Small Ubiquitin-related MOdifier (SUMO) proteins are ubiquitously expressed in eukaryotic cells (Melchior F *et al.*, 2003), and implicatedly regulate various cellular processes, e.g. stress responsing (Huang TT *et al.*, 2003), cell-cycle progression (Muller S *et al.*, 2001; Pinsky BA *et al.*, 2002; Seeler JS *et al.*, 2003), and gene expression (Muller S *et al.*, 2004), etc.

SUMO proteins belong to the superfamily of Ubiquitin-like modifiers (UBLs) (Schwartz DC *et al.*, 2003), and consist of three components in mammalian: SUMO-1, SUMO-2 and SUMO-3 (Saitoh H *et al.*, 2000). Recently, another component SUMO-4 was discovered in human (Bohren KM *et al.*, 2004). SUMO proteins are highly conserved from yeast to human.

There are only 12 experiment-verified SUMO substrates before 2000 (Melchior F, 2000). But now, > 60 SUMO substrates were discovered (Seeler JS *et al.*, 2003) and it becomes a fascinating hot area to search for new SUMO substrates in mammlians, especially in human.

Conventional experimental approaches can identify SUMO substrates, but they are tedious and time-consuming. Small-scale analysis of SUMO substrates could improve the efficiency by the method of affinity chromatography-coupled high-pressure liquid chromatography/tandem mass spectrometry (Zhao Y *et al.*, 2004); yet only 4 previously characterized and 18 novel potential SUMO substrates were found.

The majority of the SUMO substrates have a consensus motif with four amino acids. There are several motifs proposed in the literatures: such as ψ-K-X-E (Seeler JS, *et. al.* 2003, Zhao Y, *et. al.* 2004), and [VILAFP]-K-X-[EDNGP] (Puntervoll P, et. al. 2003), etc. We are mainly focusing on reducing the number of potential false positives by the phylogenetic conservation in this work. So we just choose the former one as the consensus motif. And a nuclear localization signal (NLS) suffices to produce a SUMO conjugate in vivo (Manuel S *et al.*, 2001), with only a few exceptions (Seeler JS *et al.*, 2003).

With these molecular characteristics, it's possible to predict the SUMO substrates in silico by pattern recognition, which has been widely used in the biological domain/motif predictions, e.g. the PROSITE database (Falquet L *et al.*, 2002), and the transcription factor binding sites (TFBSs) (Quandt, K *et al.*, 1995), etc. But such methods will also generate too many false positive results. We found that *rVista*, combining the pattern recognition and phylogenetic conservation approaches, has much better predicting performance of transcription factor binding sites than before (Loots GG *et al.,* 2002). To eliminate the potential false positives satisfying the above patterns, here we combined the pattern recognition and phylogenetic conservation approaches to predict the SUMO substrates and got 2,683 results in both mouse and human.

## Methods

We downloaded the orthology-relationship data of mouse and human with the corresponding sequences from the InParanoid database (Version 2.6, 30/03/2004) (Remm M *et al.,* 2001), which contains orthology information among 7 model organisms, including E.coli, S.cerevisiae, A.thaliana, C.elegans, D.melanogaster, M.musculus and H.sapiens. Here we only used the orthology information between human and mouse.

For the 34,499 mouse sequences and 36,379 human sequences in InParanoid, we firstly scanned the sequences for the consensus motif ψ(A, F, I, L, M, V, W)-K-X-E in mouse and human respectively. Sequences without such motif were excluded. Then we got 13,026 sequences in mouse and human respectively.

By PSORT II (Nakai K *et al.,* 1999), we predicted the sub-cellular localization of the retained sequences. Only proteins with predicted nuclear localization were retained. After this step, there were 6,662 sequences in mouse and 7649 in human respectively.

In order to eliminate the potential false positive results, we followed a simple rule below: for the pairwise orthologs among the retained mouse and human proteins, they must have at least one consensus SUMO substrate motif at the same position after sequence alignment. So proteins without orthologs were excluded firstly. We aligned the sequences of the remaining ortholog pairs respectively by ClustalW (Thompson JD *et al.,* 1994). After the sequence alignment, orthologs sharing no SUMO target motif at the same position were also excluded. Finally we got 2,683 proteins in both mouse and human.

## Results and Discussion

### Using homology information to eliminate the potential false positive

Since SUMO substrates and their sites in mammalians especially in human is more important and urgent than in other species recently, and the orthology information between human and mouse is more abundant and integrated, we used the phylogenetic conservation between human and mouse to reduce the potential false positives.

We got 67 experiment-verified sumoylation sites from the published work. The sequence logo of the motifs was generated using relative entropy (Gorodkin J *et al.*, 1997) (see in Figure 1). Although there are some atypical amino acid which could rarely appear in the first or fourth position, we follow this simple motif, ψ(A, F, I, L, M, V, W)-K-X-E, which holds for the majority of known sumoylation sites.

Most of the SUMO substrates are nuclear proteins, with only a few exceptions, such as the cytoplasmic proteins GLUT1, and GLUT4. So we use PSORT II to predict the sub-cellular locations of the potential proteins. The proteins without a "nuc" (nuclear) hit by PSORT II prediction were excluded.

We checked the published references related to SUMO, and 79 experiment-verified SUMO substrates were listed in TABLE 1. Our method can predict 58 of them as true positives, with accuracy ~73%. For the wrongly predicted SUMO substrates, six proteins, such as CREB and Daxx, have no consensus motif, another nine proteins don't get a "nuc" (nuclear) hit by PSORT II prediction and the other six proteins were excluded by orthology information.

By orthology information, we reduced the predicted results 7, 649 to 2, 683 in human. 64 out of 74 known SUMO substrates can be predicted correctly with only the consensus motif and the sub-celluar localization. Combining with the orthology restriction, another six true results were missed. So it can be concluded that phylogenetic conservation could be a great help to SUMO substrate prediction.

### Comparison to an existing tool SUMOplot for sumoylation sites prediction

If a SUMO substrate was predicted or verified, could we recognize the sumoylation sites correctly? In order to answer this question, we compared our method with the existing tool

SUMOplot (http://www.abgent.com/default.php?page=sumoplot), which is based on pattern recognition only.

54 verified sumoylation sites of 37 known SUMO substrates were chosen. We only considered the SUMO substrates that can be predicted correctly by our method. And the results were shown in TABLE 2.

Here we defined the sensitivity ($Sn$) and specificity ($Sp$) of the prediction as: $Sn=m/N$, and $Sp=m/M$ respectively, where $m$ is the number of predicted sumoylation sites which are true positives, $N$ is the number of experiment-verified sumoylation sites, and $M$ is the number of the predicted sites.

For total 54 known sumoylation sites, our method could recover 42 of them with the $Sn$ of ~78%, which is similar to the SUMOplot results ~78% (Motifs with high probability) or ~81% (all). Yet the specificity of our approach is significantly improved to ~57% (42 in 74 totally), comparing to SUMOplot ~28% (Motifs with high probability) or ~14% (all).

Although we missed several true results, our method successfully reduced the number of the potential false positives significantly with slightly reducing the sensitivity. For identification of SUMO substrates and sites in other organisms, we may use several available systems, such as ELM (Puntervoll P 2003).

For convenience and data security, we design the program for Windows version written in Delphi, which can be used in local.

## Acknowledgements

We thank Yi Xing (UCLA) for helpful discussions and critical reading of this manuscript. This work was supported by grants from Chinese Natural Science Foundation (39925018 and 30121001), Chinese Academy of Science (KSCX2-2-01), Chinese 973 project (2002CB713700), and American Cancer Society (RPG-99-173-01) to X. Yao. X. Yao is a GCC Distinguished Cancer Research Scholar.

**Figure 1 - The sequence logo of verified sumoylation motifs**

Curated from the published work, we got 67 experiment-verified sumoylation sites. And the sequence logo of the motifs was generated using relative entropy.

**TABLE 1 – The prediction results of the known SUMO substrates**

79 experiment-verified SUMO substrates are listed. Our method can predict 58 of them correctly (~73%). For the wrongly predicted SUMO substrates, six proteins, such as CREB and Daxx, have no consensus motif, another nine proteins don't get a "nuc" (nuclear) hit by PSORT II prediction, and the left six proteins were excluded by orthology information.

**TABLE 2 – The comparison of the sumoylation site prediction against SUMOplot**

Fifty-four verified sumoylation sites of 37 known SUMO substrates were chosen. Our method got nearly the same sensitivity as the existed tool SUMOplot, with 42 against 42 (high probability) or 44 (all) true positives respectively. But our method outperformed SUMOplot significantly by specificity, with 74 against 152 (high probability) or 324 (all) predicted sumoylation sites respectively. So our method greatly reduced the number of false positives while still kept a satisfying sensitivity.

Figure 1

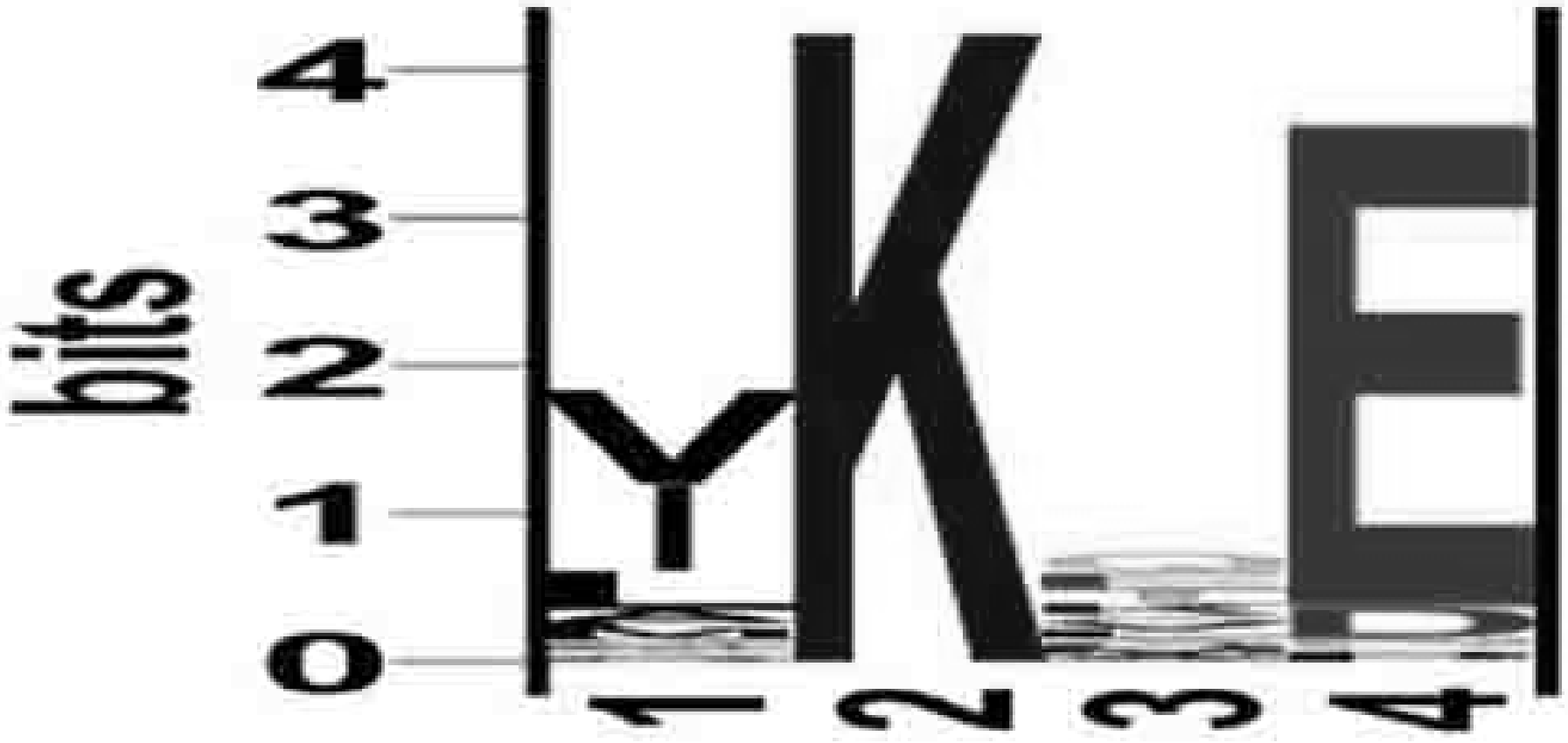

| Protein name | Sumoylation sites | References | Prediction |
|---|---|---|---|
| AP-2 alpha | K10 | 12072434 | Yes |
| AP-2 beta | K10 | 12072434 | Yes |
| AP-2 gamma | K10 | 12072434 | Yes |
| AR (androgen receptor) | K386, K520 | 11121022 | Yes |
| ARNT (aryl hydrocarbon receptor nuclear transporter) | K245 | 12354770 | Yes |
| Axin | | 12223491 | Yes |
| Bach2 | | 15060166 | Yes |
| C/EBP beta-1 | K173 | 12810706 | Yes |
| C/EBPalpha (CCAAT/enhancer-binding protein alpha) | K159 | 12511558 | Yes |
| c-Jun | K229 | 10788439 | Yes |
| c-Myb | K503, K527 | 12631292 | Yes |
| *CREB (cAMP-response element-binding protein)* | *K285, K304* | *12552083* | *No* [a] |
| *CtBP1* | *K428* | *12769861* | *No* [b] |
| *Daxx* | *K630, K631* | *12150977* | *No* [a] |
| *Dnmt3a* | | *14752048* | *No* [b] |
| *Dnmt3b* | | *14752048* | Yes |
| Dynamin -1 | | 15123615 | Yes |
| Dynamin -2 | | 15123615 | Yes |
| Dynamin -3 | | 15123615 | Yes |
| Elk-1 | K230, K249 | 14992729 | Yes |
| FAK (Focal adhesion kinase) | K152 | 14500712 | Yes |
| GATA-2 | | 12750312 | Yes |
| *GLUT1* | | *10655495* | *No* [b] |
| *GLUT4* | | *11842083* | *No* [b] |
| GR (glucocorticoid receptor) | K277, K293 | 12144530 | Yes |
| GRIP1 | K239, K731, and K788 | 12060666 | Yes |
| *HDAC1* | *K444, K476* | *11960997* | *No* [b] |
| *HDAC4* | *K559* | *12032081* | *No* [c] |
| HIPK2 | K1182 | 10535925 | Yes |
| *histone H4* | | *14578449* | *No* [a] |
| hnRNP C | K237 | 15082759 | Yes |
| *hnRNP M* | | *15082759* | *No* [b] |
| HSF1 (Heat shock transcription factor 1) | K298 | 11514557 | Yes |
| HSF2 (Heat shock transcription factor 2) | K82 | 11278381 | Yes |
| IkappaBalpha | K21 | 9734360 | Yes |
| *IRF-1 (Interferon regulatory factor-1)* | | *12387893* | *No* [a] |
| LEF1 | K27, K269 | 11731474 | Yes |
| Mdm2 | | 11384992 | Yes |
| MR (mineralocorticoid receptor) | | 14500761 | Yes |
| NEMO/IKKgamma | K277, K309 | 14651848 | Yes |
| NFAT1 | | 15117942 | Yes |
| Nurr1 (NR4A2, RNR-1,TINUR, HZF-3) | K91, K577 | 14559918 | Yes |
| p300/CBP | K1017, K1029 | 12718889 | Yes |
| *p53* | *K386* | *10788439* | *No* [c] |
| p73alpha | K627 | 10961991 | Yes |
| PC2 | | 12679040 | Yes |
| Pdx1 (Pancreatic duodenal homeobox-1) | | 12488243 | Yes |
| *PIAS1* | | *12077349* | *No* [c] |
| *PIASx-beta* | | *12077349* | *No* [a] |
| *PLZF (Promyelocytic leukemia zinc finger)* | *K242* | *14527952* | *No* [c] |
| PML (promyelocytic leukaemia protein) | K65, K160, and K490 | 10525530 | Yes |
| PPAR-gamma | | 15123625 | *No* [c] |
| PR (progesterone receptor) | K388 | 12529333 | Yes |
| RanBP2/NUP358 | | 15037602 | Yes |
| *RanGAP1* | *K526* | *9442102* | *No* [b] |
| SALL1 | K1086 | 12200128 | Yes |
| SATB2 | | 14701874 | Yes |
| SENP1 | | 14563852 | Yes |
| Smad4 | K113, K159 | 12621041 | Yes |
| Sox6 | | A | Yes |
| Sox9 | | A | Yes |
| *Sp100* | *K297* | *10212234* | *No* [c] |
| Sp3 | K539 | 12419227 | Yes |
| SREBP-1a | K123, K418 | 12615929 | Yes |
| *SREBP-2* | *K464* | *12615929* | *No* [b] |
| SRF (serum response factor) | K147 | 12788062 | Yes |
| *STAT1* | *K703* | *12764129* | *No* [b] |
| steroid receptor coactivator SRC-1/NCoA-1 | K732, K774 | 12529333 | Yes |
| Tcf-4 | K297 | 12727872 | Yes |
| TDG | K330 | 11889051 | Yes |
| TEL | K99 | 12626745 | Yes |
| TFII-I | | 15016812 | Yes |
| TIF1alpha | K690, K708 | 11313457 | Yes |
| TOPO I | K117, K153 | 12439742 | Yes |
| topoisomerase II alpha | | 14597774 | Yes |
| topoisomerase II beta | | 12832072 | Yes |
| Topors | K560 | 14516784 | Yes |
| WRN | | 10806190 | Yes |
| *zinc finger protein APA-1* | | *12370286* | *No* [a] |

[A] *R. Fernandez-Lloris, F. Ventura1 and R. T. Hay. (2002) Post-translational Sox6 protein modification by SUMO-1. 28th Meeting of the Federation of European Biochemical Societies 20-25, Istanbul, Turkey*

[a] *No consensus motif (6 proteins)*

[b] *Not "nuc" (nuclear) hit by PSORT II prediction (9 proteins)*

[c] *Excluded by orthlogy information (6 proteins)*

**TABLE 1**

| Protein name | Sumoylation sites | | |
|---|---|---|---|
| | Verified | SSP 1.0 | SUMOplot |
| AP-2 alpha | K10 | **IKYE**[a] | 1/1[b]; (1/4)[c] |
| AP-2 beta | K10 | **IKYE** | 1/1; (1/5) |
| AP-2 gamma | K10 | **IKYE** | 1/1; (1/4) |
| AR (androgen receptor) | K386, K520 | **IKLE, VKSE** | 2/3; (2/6) |
| ARNT (aryl hydrocarbon receptor nuclear transporter) | K245 | **VKKE** | 0/2; (0/5) |
| C/EBP beta-1 | K173 | **LKAE** | 1/2; (1/4) |
| C/EBPalpha (CCAAT/enhancer-binding protein alpha) | K159 | **LKAE** | 1/1; (1/1) |
| c-Jun | K229 | AKME, **LKEE**, IKAE | 1/3; (1/4) |
| c-Myb | K503, K527 | **IKQE, IKQE** | 2/5; (2/10) |
| Elk-1 | K230, K249 | **VKVE** | 2/2; (2/4) |
| FAK (Focal adhesion kinase) | K152 | WKYE | 1/5; (1/17) |
| GR (glucocorticoid receptor) | K277, K293 | **VKTE, IKQE**, VKRE | 0/1; (0/5) |
| GRIP1 | K239, K731, K788 | **VKLE, MKQE** | 2/7; (2/17) |
| HIPK2 | K1182 | LKIE, LKPE | 0/3; (0/7) |
| hnRNP C | K237 | IKKE, **VKME** | 0/4; (1/12) |
| HSF1 (Heat shock transcription factor 1) | K298 | VKPE, LKSE, MKHE, **VKEE** | 1/6; (1/9) |
| HSF2 (Heat shock transcription factor 2) | K82 | **VKQE**, IKQE, LKSE | 1/6; (1/9) |
| IkappaBalpha | K21 | **LKKE**, MKDE | 1/4; (1/4) |
| LEF1 | K27, K269 | **FKDE, VKQE** | 2/3; (2/7) |
| NEMO/IKKgamma | K277, K309 | **AKQE**, LKEE | 1/8; (1/13) |
| Nurr1 (NR4A2, RNR-1,TINUR, HZF-3) | K91, K577 | **IKVE, LKLE** | 2/4; (2/10) |
| p300/CBP | K1017, K1029 | MKTE, VKEE, VKVE, VKEE, FKPE | 2/11; (2/22) |
| p73alpha | K627 | **IKEE** | 1/3; (1/7) |
| PML (promyelocytic leukaemia protein) | K65, K160, K490 | **LKHE, IKME** | 3/6; (3/7) |
| PR (progesterone receptor) | K388 | **IKEE** | 1/3; (1/6) |
| SALL1 | K1086 | IKTE, **IKTE** | 1/7; (1/15) |
| Smad4 | K113, K159 | **VKDE** | 1/3; (1/7) |
| Sp3 | K539 | IKDE, **IKEE** | 1/3; (1/5) |
| SREBP-1a | K123, K418 | **IKEE**, LKQE, **VKTE** | 2/7; (2/11) |
| SRF (serum response factor) | K147 | **IKME** | 1/1; (1/3) |
| steroid receptor coactivator SRC-1/ NCoA-1 | K732, K774 | AKAE, **IKLE, VKVE**, IKLE, IKSE | 1/7; (1/13) |
| Tcf-4 | K297 | FKDE, **VKQE** | 1/6; (1/10) |
| TDG | K330 | **VKEE** | 0/0; (0/8) |
| TEL | K99 | IKQE | 0/2; (0/3) |
| TIF1alpha | K690, K708 | **IKQE, VKQE**, IKLE | 2/5; (2/11) |
| TOPO I | K117, K153 | IKKE, **IKTE**, IKEE, FKIE, IKGE, MKLE | 2/12; (2/29) |
| Topors | K560 | LKRE | 0/4; (1/10) |
| ***Total sites*** | 54 | 42/74[d] | 42/152; (44/324) |

[a] *SSP 1.0 hits are in bold character font*

[b] *SUMOplot hits (Motifs with high probability) / total predicted sites (Motifs with high probability)*

[c] *SUMOplot hits (all) / total predicted sites (all)*

[d] *SSP 1.0 hits/total predicted sites*

# TABLE 2